\documentclass[11pt,aps,prd,floatfix]{revtex4}
\usepackage{bm}
\usepackage{latexsym}
\usepackage{dcolumn}
\usepackage{amsmath,amsfonts,amssymb}
\usepackage{graphicx,epsfig,float}
\usepackage{verbatim}
\usepackage{color}
\def\p {\partial}

\def\be {\begin{equation}}
\def\ee  {\end{equation}}
\def\bea {\begin{eqnarray}} 
\def\eea {\end{eqnarray}}

\reversemarginpar
\begin{document}
\title{Does quantum gravity relate the constants of nature?}
\author{Viqar Husain}
\email{vhusain@unb.ca}
\author{Suprit Singh}
\email{suprit.singh@unb.ca}

 \affiliation{Department of Mathematics and Statistics, University
of New Brunswick, Fredericton, NB, Canada E3B 5A3} \pacs{...}

\begin{abstract}
\smallskip

The central equation of quantum gravity is the Wheeler-DeWitt equation. We give an argument suggesting that  exact solutions of this equation give a surface in the space of coupling constants. This provides a mechanism for determining the cosmological constant as a function of the gravitational and other interaction constants.  We demonstrate the idea by computing one such surface in a cosmological model. 

\bigskip

\end{abstract}
\date{\today}
\maketitle
\vskip 1 cm
\noindent
\maketitle

%%%%%%%%%%%%%%%%%%%%%%%%%%%%%%%%%%%%%%%%%%%%%%%%%%%%%%%%%%%%%%%%%%%%%%%%%%%%%%%%%%%%%%%%%%%%%
   %\vfill
   %\eject 

\section{Introduction}     

One of the approaches to quantum gravity is the canonical quantization program. This originated in a now classic paper by DeWitt  \cite{DeWitt:1967yk}. The basic idea of the proposal is to impose the classical constraints of general relativity as operator conditions on wavefunctionals of the spatial metric. Schematically, this is quantified as
\be
\hat{{\cal H}}(q,\delta/\delta q)\  \Psi[q] =0, \label{wdw}
\ee
an equation known as the Wheeler-DeWitt (WDW) equation. Along with this equation is one that imposes invariance of the state under spatial diffeomorphisms. 

There is by now a vast literature on this proposal. This covers the homogeneous cosmological models, where this equation simplifies to a condition in quantum mechanics \cite{Ryan:1975jw}, to more complex systems with local dynamical degrees of freedom \cite{FernandoBarbero:2010qy}. There remain however significant difficulties in both formulating the equation precisely, and in finding and interpreting meaningful solutions. Much of the work on this equation is of a formal nature \cite{Hartle:1983ai}, since both operator ordering and regulation are serious concerns \cite{Tsamis:1987wf}. More recent works in the loop quantum gravity formalism have yielded a useful Hilbert space in which these problems can at least be carefully studied \cite{Thiemann:2007zz}.   

The WDW equation acquires additional terms when matter and a cosmological constant are added. The device usually employed in  attempts to solve the extended equation is a product ansatz, $\Psi[q,\phi]= \phi[q]\chi[q,\phi]$, where $\phi$ denotes the matter fields. The idea is to use a WKB type of methodology to find approximate solutions. This leads to a formal derivation of the theory of quantum fields in curved spacetime by treating gravity as a ``slow" variable and``matter" as the fast one (see eg. \cite{Kiefer:2018xfw} and references therein). 
 
In this general setting the  WDW operator (\ref{wdw}) depends on the gravitational and cosmological constants $G$ and $\Lambda$, and  several matter coupling constants $\mu$. At present it is far from clear if exact solutions of the equation can be found for arbitrary values of all these parameters in any but the simplest of cases. Indeed, in a recent work  \cite{Dittrich:2016hvj} it is suggested that even simple versions of WDW equations may have no semiclassical limit, and no solutions, if standard quantization methods are used; the suggested solution to this dilemma is to use a Hilbert space with an uncountable basis so there is more ``room" to solve the equation.
  
In this note we propose an idea that ensures multiple exact solutions of the WDW equation within the standard quantization procedure. The idea is to solve the eigenvalue problem for the WDW operator, and set a selected eigenvalue to be zero. This gives a relation between coupling constants. We demonstrate a proof of principle by implementing the  WDW operator in a model system of a Friedmann-Robertson-Walker (FRW) cosmology, and argue that our method applies generally. We discuss consequences of this approach for questions such as singularity avoidance and matter-gravity entanglement in quantum gravity.
 
\section{Wheeler-DeWitt eigenvalue problem}
 
The main feature of the gravitational Hamiltonian density that makes it very different from conventional systems is that the gravitational kinetic term contains a combination of the configuration and the momentum variables. This is seen for example in the Arnowitt-Deser-Misner (ADM) formalism for general relativity, and it is also apparent in connection-triad Ashtekar-Barbero variables. With phase space variables $(q_{ab},\pi^{ab})$ specifying the spatial metric and its conjugate momentum, the gravitational part of the Hamiltonian density is 
\bea
{\cal H}^G_{\text{kin}} \equiv \frac{1}{\sqrt{q}} \left( \pi^{ab}\pi_{ab} - \frac{1}{2} \pi^2 \right). 
\eea
The matter contributions to the WDW operator containing couplings to the spatial metric. For a scalar field with potential $V$, the matter Hamiltonian is 
\bea
{\cal H}^M \equiv \frac{1}{2}\left( \frac{p_\phi^2}{\sqrt{q}} + \sqrt{q} q^{ab}  \p_a\phi\p_b\phi \right) + \sqrt{q}\,V(\phi;\mu).
\eea

Like any quantum system, the spectrum of the  WDW operator will be a function of all the constants appearing in it, that is, 
\bea
\left[ {\cal H}^G_{\text{kin}} - \sqrt{q}(R(q) - \Lambda) + G {\cal H}^M   \right] \Psi = E(G,\Lambda, \mu)  \Psi,
\eea
where $R(q)$ is the Ricci scalar of $q_{ab}$, $G$ and $\Lambda$ are the gravitational and cosmological constants, and $\mu$ is a matter coupling constant.  Exact physical states of quantum gravity are those for which $E(G,\Lambda, \mu) = 0$. If it turns out that there are no solutions for arbitrary values of these parameters, then the only possibility is to take (selected)  non-zero eigenvalues and set them to zero. This would lead to relations of the form $\Lambda = f(G, \mu)$. We will see below that this becomes necessary if the WDW operator is truncated to a finite dimensional matrix, as must happen for calculations   in discrete models.

The second eigenvalue problem comes from the spatial diffeomorphism operator, namely 
\be
\left(- D_c\pi^{c}_{\ a}  + G p_\phi \p_a \phi\right) \Psi = C(G) \Psi. 
\ee 
The eigenvalue for this second equation is not a function of $\Lambda$ or any other matter coupling constants, since these parameters do not appear on the left hand side. Therefore the physical state condition is either the root condition $C(G)=0$  that  selects values of $G$, or that the chosen eigenvector  from solving the first equation automatically solves this condition as well.  In either case both conditions would have be solved simultaneously. 

As a first step we demonstrate this basic idea on an oft-studied system, the homogeneous and isotropic cosmology.  It has the advantage that the spatial diffeomorphism constraint is classically not present, and so no corresponding quantum condition is necessary, at least for the usual path of reduction before quantization. 

\section{A model system: cosmology}

We consider the reduction to FRW cosmology where the gravitational phase space  variables are $(x, p)$, with $x=a^3$, (where $a$ is the scale factor), and the matter variables are $(\phi,p_\phi)$.  The hamiltonian constraint in these variables is
\be
{\cal H} = - ( p^2x  + k x^{1/3} - \Lambda x ) + G\left( \frac{p_\phi^2}{x} + x V(\phi) \right),
\ee 
where $k$ is the spatial curvature. 

The Hilbert space for quantization  is the tensor product  $H_G\otimes H_M$ of the matter and gravity sectors, and the the corresponding WDW operator is formally the expression
\bea
 \hat{{\cal H}} =& - \left( \widehat{p x p}  + k \hat{x}^{\frac{1}{3}}- \Lambda 
 \hat{x}  \right)  \otimes \hat{I}_M \nonumber\\
 &+\, G\left(\widehat{\frac{1} {x}}  \otimes \hat{p}_\phi^2 + \mu\ \hat{x}\otimes \hat{\phi}^2 \right),
\label{FRW}
\eea 
where $\hat{I}_M$ is the identity in $H_M$.   

A direct approach for finding the spectrum of this operator is to use the occupation number basis to write all operators in terms of $\hat{a}$ and $\hat{a}^\dagger$, that is using $x = (\hat{a}^\dagger + \hat{a})$ and $p = i(\hat{a}^\dagger - \hat{a})$, and similarly for the matter variables. This expectedly yields an infinite dimensional matrix. We  consider two choices for the matter sector: (i) as a ``particle with spin" using the Pauli kinetic operator $(p_\phi \cdot \sigma_y)^2$, where $\sigma_y$ is a Pauli matrix, so the matter Hilbert space is  also a tensor product  of the scalar and spin one-half components, i.e. $H_M = L^2(\mathbb{R}) \times H_{\frac{1}{2}}$; and (ii) $H_M= L^2(\mathbb{R})$, which is the usual case. 

It is possible to get an idea of the spectrum by truncating the WDW matrix, which is equivalent to limiting the excitation level in the oscillator basis; this may be viewed as a ``cutoff."  While this may seem artificial, similar uses of cutoffs is inherent in path integral approaches to quantum gravity that are based on triangulations of the manifold;  the amplitudes so obtained are expected to be physical states, and therefore implicitly solutions of a truncated WDW operator.   
 
 We now note that in any such truncation it is readily shown numerically that zero is not in the spectrum.  Therefore the only other way to solve (\ref{FRW}), with no other approximations and a given value of $k$, is to impose 
\bea
E(G, \Lambda,\mu) =0    
\eea  
for a chosen eigenstate.  This yields a solution surface $\Lambda = \Lambda (G,\mu)$ for the corresponding eigenvalue. Each level of truncation, and each eigenvalue within a truncation, would yield a different curve. In principle this approach, although unorthodox, leaves open the possibility of finding solution curves that match the currently observed values of $G$ and $\Lambda$. Another interesting feature of this mechanism for solving the WdW equation is that zero eigenstates  are such that the matter and gravity subsystems may be entangled for sufficiently large truncations (as we show below).

Let us demonstrate how this approach works with two particular truncations, which although small, gives the basic idea. For the Pauli particle case we consider the truncation of the WDW operator (\ref{FRW}) where the gravity and matter sectors are respectively  $2\times 2$  and $8\times 8$ matrices. That is, we truncate $p_\phi$ and $\phi$ to $4\times4$ matrices, which gives the $8\times 8$ system after a tensor product with  the Pauli matrix $\sigma_y$.  The resulting $16\times 16$ matrix representing the WdW operator has rank 16. 

For the scalar particle (i.e. no spin), we take gravity and matter sectors to be both $4\times4$ matrices. Again we find that that the rank of the resulting matrix is 16. It is readily shown that this pattern generalizes to higher dimensional truncations.

As a result, the only way to find zero eigenvalues is by choosing $G$, $\Lambda$ and $\mu$  to be functionally dependent in a specific way. Such relations may be found by diagonalizing the matrix and setting an eigenvalue to zero. (This procedure is readily automated  using a package such as MAPLE.)

\begin{figure}[t]
\centering
\includegraphics[scale=0.95]{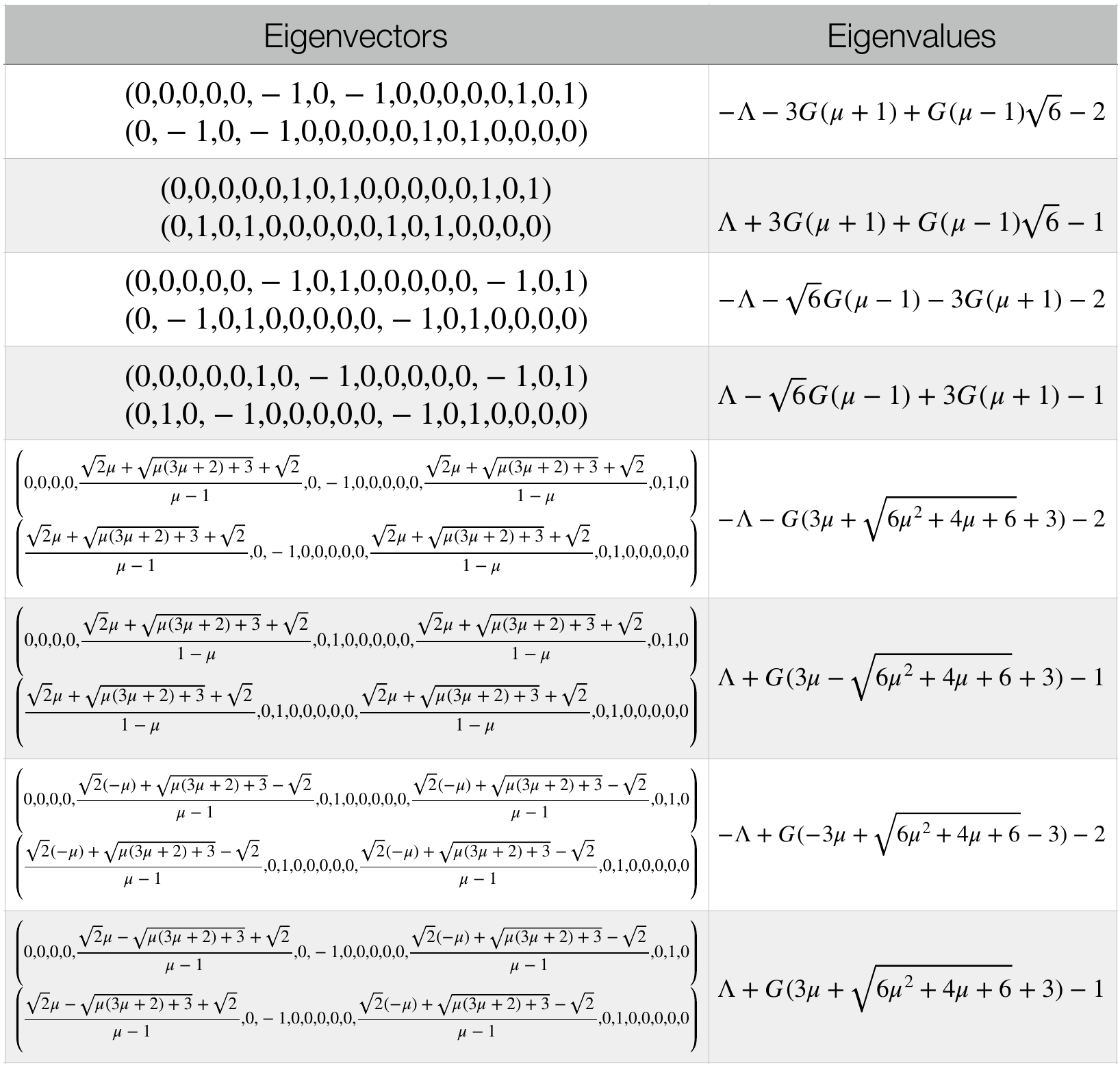}
\caption{Eigensystem for the model Hamiltonian with $k =2$. There are eight doubly degenerate eigenvalues which are linear in $\Lambda$. The eigenvectors are correct up to an overall normalization factor.}
\label{table}
\end{figure}

For the scalar-spin case,  with $k=2$, there are eight doubly degenerate eigenvalues. These are shown in  Fig. \ref{table}. We note that all the eigenvalues are linear in $\Lambda$, a feature that  is not surprising because (i) $\Lambda$ appears linearly in the WDW operator, and (ii) there no terms where it appears in combination with $G$ or $\mu$. Furthermore, in this truncation all the eigenvectors turn out to be  product states, and so have  zero entanglement.  We will see below that this is not necessarily the case  in our next example, where the gravity truncation is not this small.

\begin{figure}[H]
\centering
\includegraphics[scale=0.5]{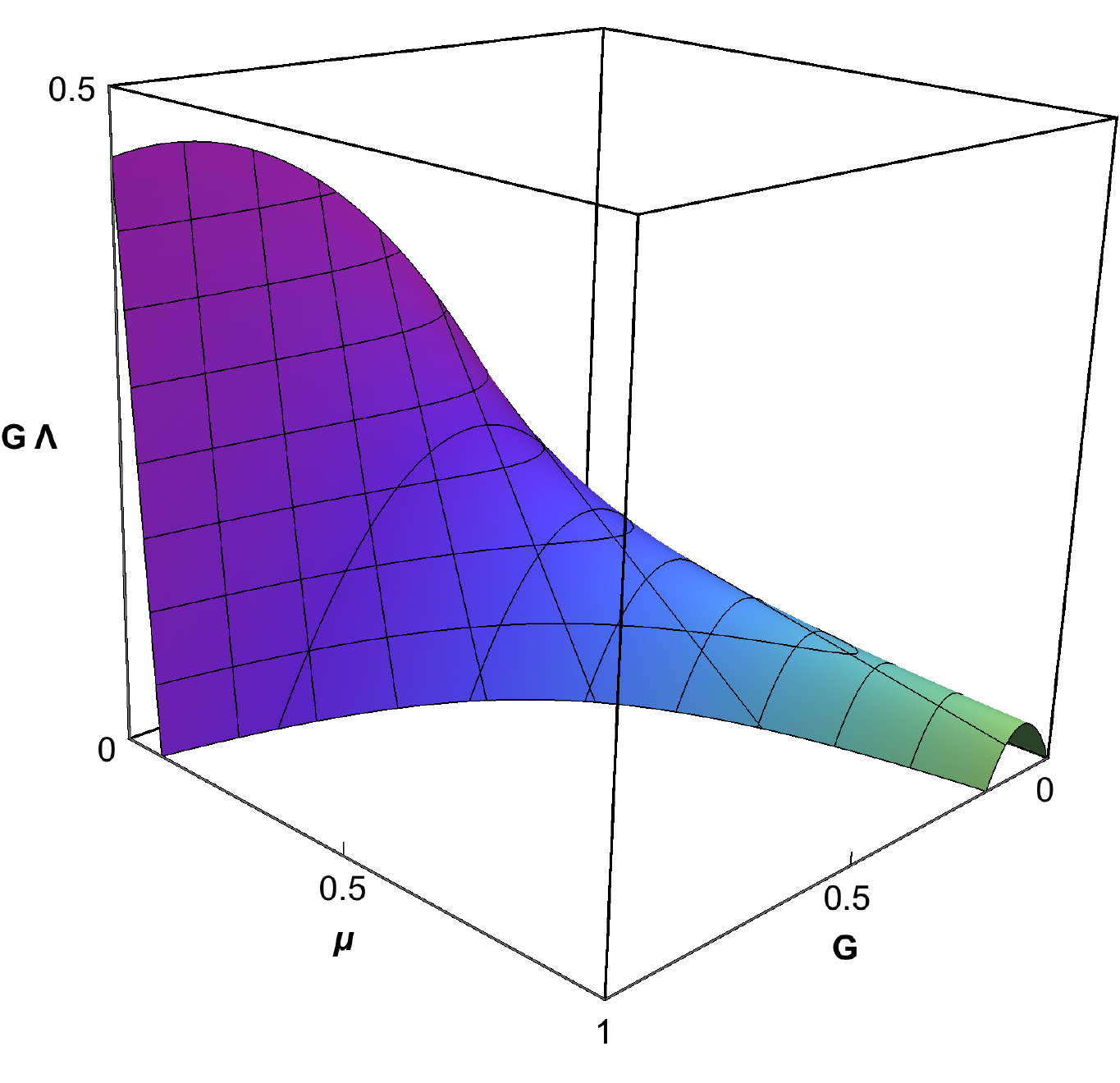}\quad\qquad\includegraphics[scale=0.5]{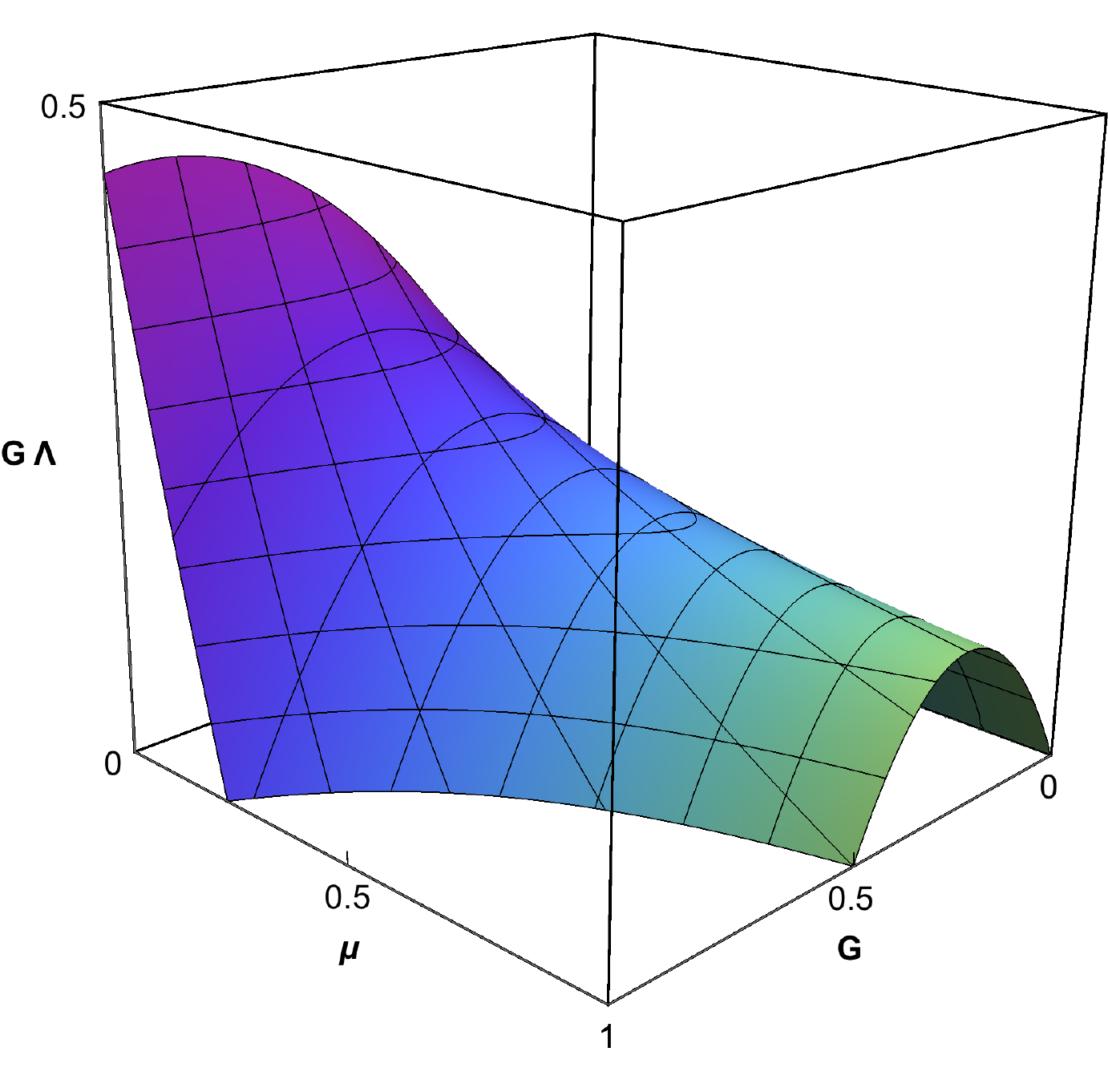}
\caption{Solution surfaces for the truncated Wheeler-DeWitt equation (\ref{FRW}) obtained by setting the second and sixth eigenvalues to zero respectively. It is evident that there are solutions for $\Lambda/G$ near zero for positive mass $\mu$ in geometrized units.}
\label{LGmu}
\end{figure}
 
Zero eigenvalue surfaces are obtained by selecting any eigenvalue from the system and setting it to zero. For example, the second and sixth eigenvalues from the above set give the surfaces
\begin{align}
\Lambda = & 1 + G\sqrt{2} \sqrt{3  \mu ^2+2  \mu +3 } - 3 G (\mu + 1)\nonumber\\
\Lambda = & 1+ G \left(-\left(\sqrt{6}+3\right) \mu +\sqrt{6}-3\right) 
\end{align}
This is shown in Fig. \ref{LGmu}; it demonstrates that it is possible to obtain a very small value of $\Lambda$  for order unity values of  $G$ and $\mu$. What about the other eigenvalues? Similar plots show that  only a few from this list give solution surfaces where $G$ and $\mu$ are positive.   

We now turn to our next example, where both gravity and matter are truncated to $4\times 4$ and the resulting WDW operator is $16\times 16$. In this case it is not possible to extract analytical expressions for the WDW eigenvalues. Instead we computed the spectrum numerically 
for $G=1$ and with $\Lambda, \mu \in [0,2]$.  Of the sixteen eigenvalues $E(G=1,\Lambda,\mu)$, only two have zeros for certain ranges of $\mu$ and $\Lambda$. Examples of such surfaces are shown in Fig. 3. Both show that it is possible to have very small values of $\Lambda$ for 
order one values of $G$ and $\mu$. 

\begin{figure}[H]
\centering
\includegraphics[scale=0.5]{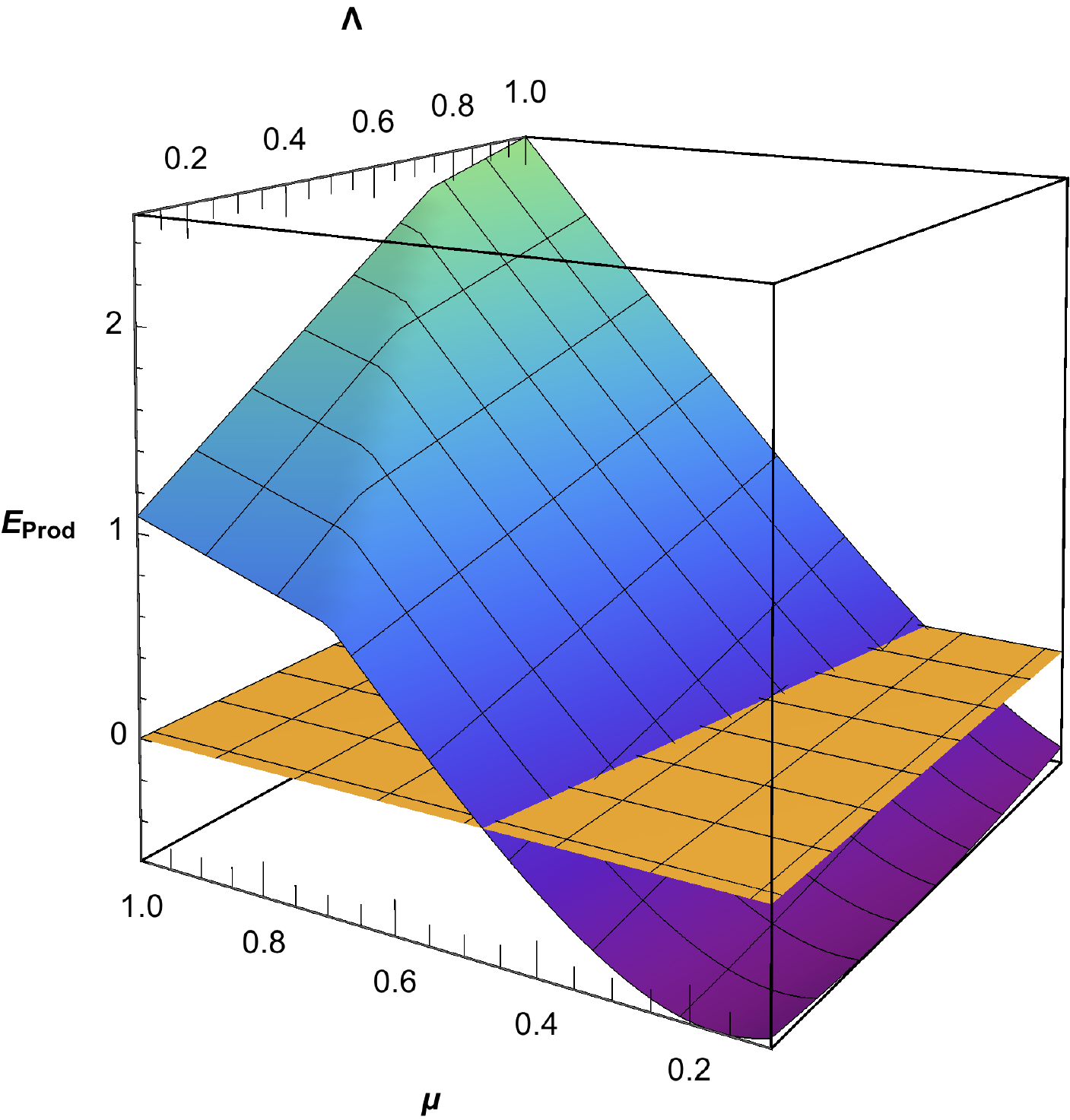}  \quad\qquad\includegraphics[scale=0.5]{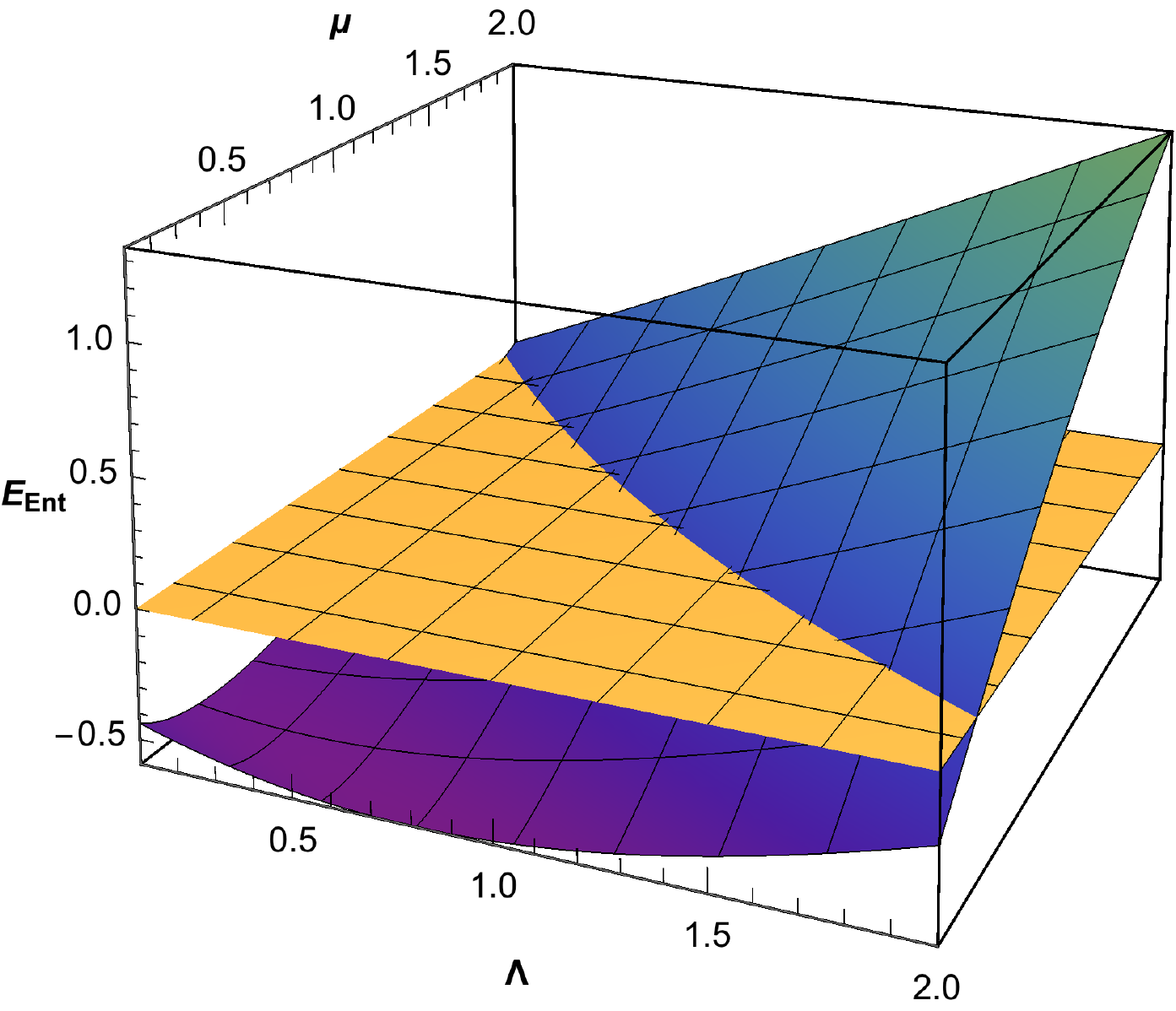}
\caption{Energy surfaces for a two representative eigenvalue surface that cut the zero plane with $G=1$. The intersection is the WDW solution curve. All states in the left frame are product states, and those on the right are entangled.}
\label{LGmu2}
\end{figure}

The Von-Neuman entanglement entropies $S$ for the corresponding eigenvectors (which also depend on $\Lambda$ and $\mu$) can be calculated  from the density matrix $\rho_M = {\text Tr}_G  (\rho)$. This may be done  for  selected values $\Lambda,\mu$ on the curve
where the surface intersects the zero plane. For the zero eigenvalue curve  in the left frame of Fig. 3 we find that all entanglement entropies are zero, whereas for the right frame all states on the curve are entangled. Two separated points on the latter indicate the degree of entanglement.

One of the tests of any quantum gravity model is to see if it predicts singularity avoidance. This is a feature we can test in our approach.
Take any eigenvalue of the WDW operator in our truncated system, and consider the question of the expectation value of the curvature term in the operator (\ref{FRW}).  Calculation of this expectation value in the normalized states corresponding to the second and sixth eigenvalues, corresponding  to the surfaces in Fig. 1, gives the result
\be
  \langle \psi |  \widehat{\sqrt{q} R(q)}| \psi  \rangle= \langle \psi | \hat{x}^{\frac{1}{3}} \otimes I_M| \psi  \rangle = 1.
\ee
This conforms to the fact that the eigenvectors are states of constant curvature which we expect for the FRW system. It is an indication that curvatures are bounded in the quantum theory. Similar results hold for the other eigenvectors. 
 
 \section{Discussion}
 
 We have described a new approach to solving the WDW equation, whereby the spectrum of the operator is found, and one or more eigenvalues are set to zero to obtain a relation for the cosmological constant as a function of all other coupling constants. Although unorthodox, the approach may be the only possibility for solving the WDW equation for truncated systems arising from underlying discretization. In our approach the truncation is a result of limiting excitation level in an oscillator basis. We demonstrated the feasibility of the approach by applying it to FRW cosmology coupled to a (fermionic) particle. This provides a ``proof of concept." In this model we also showed that on solutions the expectation value of curvature is generically bounded.  
  
Extensions of this idea to larger gravitational systems is clearly feasible and potentially useful. One such case is again in the cosmological setting, but with inhomogeneities introduced by perturbations. This would  give a  Hamiltonian constraint with additional terms labelled by modes, up to some maximum number.  The resulting WDW matrix  would have two truncations, one for the number of wave modes (UV cutoff), and the other for the occupation number for each mode. From the arguments we have presented, such a truncated matrix would not have a null space for arbitrary values of the couplings constant. This hypothesis may be checked numerically within a specific truncation.

It is clear that not all solutions of the WDW operator are physically meaningful, whether it is in the new approach we describe here, or conventional ones where no relation between coupling constants  is imposed. Indeed given a set of physically meaningful states, it is always possible to construct linear combinations which have no phenomenological interpretation. 

It is interesting to speculate on a curious interplay of this idea with Wilson's theory space.  The renormalization group (RG) in the truncated space of $n$ coupling constants  yields $n$ coupled differential equations for the flow of these constants under changes of scale. $G$ and $\Lambda$ are among these constants in effective field theory on a fixed background. Ultraviolet surfaces in the theory space are determined by this flow where the (``irrelevant") couplings flow to zero.  Our suggestion for solving the WDW equation (within a given truncation) would provide a surface in theory space, and this would be the constrained arena for effective field theory.
 
Another comparison of this idea is  with the collection of intuitions that underlie  ``Mach's Principle." There are many statements under this umbrella, several of which may be summarized as ``cosmic conditions affect local physics" \cite{Bondi:1996md}. A specific one is that  in general relativity,  the Hamiltonian constraint may be viewed as a manifestation of Mach's principle, because it is an elliptic equation  \cite{Barbour:1995iu}.  Our suggestion for solving the WDW equation may therefore be viewed as providing a ``Mach-like principle in the space of physical constants," whereby the cosmological constant is determined by the other constants of Nature. This would be a curious generalization of Machian ideas to quantum gravity. 

Lastly we note that there is another method in an ``emergent" gravity formalism where it is possible to derive a relationship between the cosmological constant and energy density scales in FRW cosmology \cite{Padmanabhan:2017qvh,Padmanabhan:2014nca}; at present this approach  does not appear to be related to the Wheeler-DeWitt equation in its implementation, but rather to the classical Friedmann equations with an additional assumption about cosmic information and holography. 

   \vskip 1cm 
   
\noindent{\bf Acknowledgements}  This work was supported in part by the Natural Science  and Engineering Research Council of Canada. It is an expanded version of an  essay written for the Gravity Research Foundation.

\bibliography{spindewitt-ref}

\end{document}